%% file: paper.tex
\documentclass[10pt]{llncs}

\usepackage[title]{appendix}
\usepackage{amsfonts}
\usepackage{amssymb}
\usepackage{graphicx}
\usepackage{url}
\usepackage{color}
\usepackage{tcolorbox}
\usepackage{hyperref}
\usepackage{lmodern}
\usepackage{algorithm,algpseudocode,multicol}
\usepackage{indentfirst}
\usepackage{xspace}
\usepackage{paralist}
\usepackage{wrapfig}
\usepackage{subfig}
\usepackage{mathptmx}

\usepackage{fancyhdr}
\fancypagestyle{specialfooter}{
  \fancyhf{}
  
  \fancyfoot[R]{To be published in the 24th edition of ESORICS (2019)}
}

\let\llncssubparagraph\subparagraph
\let\subparagraph\paragraph
\usepackage[compact]{titlesec}
\let\subparagraph\llncssubparagraph
\usepackage[inline]{enumitem}
\setlist[enumerate,1]{label=\textit{\alph*)}}
\usepackage[font={small,it},skip=3pt,belowskip=-7pt]{caption}

\renewcommand{\tablename}{Tab.}
\renewcommand{\figurename}{Fig.}

\algnewcommand\algorithmicassert{\texttt{assert}}
\algnewcommand\Assert[1]{\State \algorithmicassert(#1)}
\algrenewcommand\alglinenumber[1]{\scriptsize #1:}

\usepackage{listings}
\definecolor{backcolour}{rgb}{0.95,0.95,0.92}
\lstdefinestyle{lststyle}{
    basicstyle=\scriptsize,
    breakatwhitespace=false,
    breaklines=true,
    captionpos=b,
    keepspaces=true,
    numbersep=5pt,
    showspaces=false,
    showstringspaces=false,
    showtabs=false,
    tabsize=2
}

\newcommand{\name}{PDFS\xspace}
\newcommand{\teds}{TDS\xspace}

\begin{document}

\title{\name: Practical Data Feed Service for Smart Contracts}

\author{
    Juan Guarnizo \and 
    Pawel Szalachowski
}
\institute{
    Singapore University of Technology and Design\\
    \email{juan\_guarnizo@mymail.sutd.edu.sg}\\
    \email{pawel@sutd.edu.sg}
}

\maketitle

\begin{abstract}
    Smart contracts allow untrusting parties to arrange agreements encoded as 
    code deployed on a blockchain platform. To release 
    their potential, it is necessary to connect the contracts with the outside 
    world, such that they can understand and use information from other 
    infrastructures. However, there are many challenges associated with 
    realizing such a system, and despite the existence of many proposals, no 
    solution is secure, provides easily-parsable data, introduces small 
    overheads, and is easy to deploy.

    In this paper, we propose Practical Data Feed Service (\name), a system that
    combines the advantages of the previous schemes and introduces new
    functionalities. \name extends content providers by including new features
    for data transparency and consistency validations. This combination 
    provides multiple benefits like content which is easy to parse and 
    efficient authenticity verification without breaking natural trust chains.
    \name keeps content providers auditable and mitigates their malicious 
    activities (like data modification or censorship) and allows them to create
    a new business model. We show how \name is integrated with content 
    providers, report on a \name implementation and present results from 
    conducted experimental evaluations.

\end{abstract}

\thispagestyle{specialfooter}

\section{Introduction}
\label{sec:intro}
\input{sec/intro}

\section{Background}
\label{sec:pre}
\input{sec/pre}

\section{Architecture Overview}
\label{sec:overview}
\input{sec/overview}

\section{Details}
\label{sec:protocol}
\input{sec/details}

\section{Security Discussion}
\label{sec:analysis}
\input{sec/analysis}

\section{Realization in Practice}
\label{sec:implementation}
\input{sec/implementation}

\section{Related Work}
\label{sec:related}
\input{sec/related}

\section{Conclusions}
\label{sec:conclusions}
\input{sec/conclusions}

\section{Acknowledgment}
This research was supported by ST Electronics and National Research Foundation
(NRF), Prime Minister's Office Singapore, under Corporate Laboratory @
University Scheme (Programme Title: STEE Infosec - SUTD Corporate Laboratory).

\bibliographystyle{splncs04}
\bibliography{ref}

\begin{subappendices}
\renewcommand{\thesection}{\Alph{section}}
\label{sec:appendix}
\input{sec/appendix}
\end{subappendices}

\end{document}

%% file: sec/intro.tex
The concept of smart contracts was introduced by 
Szabo~\cite{szabo1996smart,szabo1997formalizing,szabo2002formal}. They allow
mutually untrusting parties to arrange and execute agreements without involving
any third trusted party. These agreements are expressed in a programming
language, hence can encode any processing logic possible to express in the used
language in a precise and unambiguous way. The concept has been unexplored for
decades; however, with the rise of Bitcoin~\cite{nakamoto2008bitcoin},
distributed consensus, and blockchain platforms in general, smart contracts can
finally be implemented in a practical way. Smart contracts deployed solely on a 
blockchain platform have some fundamental limitations. One problem is that a 
smart contract can only use resources available on the blockchain. This issue 
limits them from using external data provided by other infrastructures, like 
HTTP(S) data feeds. Ideally, smart contracts could process data provided by
other infrastructures and use that to encode processing logic. Unfortunately, 
there are many challenges associated with that.

One such challenge is the authenticity of data feeds. Data provided to a smart
contract should be authentic, so that the smart contract can verify its origin
and execute accordingly. Unfortunately, the widely deployed Transport Layer
Security (TLS) protocol~\cite{dierks2008transport} is inoperable in such a
setting. Secure web servers that deploy it (i.e., running HTTP over TLS --
HTTPS), cannot provide data authenticity to third parties like smart contracts.
First approaches to make this data accessible to smart contracts were
centralized oracles~\cite{zhang2016town,TLSnotary,ellisdecentralized,oraclize}. 
This introduced new trusted third parties which fetch HTTPS websites,
parse them, and provide the data to smart contracts (which finally process it).
These solutions present strong trust assumptions (i.e., a new trusted party).
To relax it, a concept of oracles based on trust computing was 
proposed~\cite{zhang2016town}. These oracles work similarly, however, the code
run by them is executed with the Intel's Software Guard Extensions
(SGX)~\cite{costan2016intel} framework, which allows proving attestation of the
code executed by the oracles. A disadvantage of this approach is to position
Intel as a centralized trusted entity, and SGX as a trusted technology. In
contrast to these approaches, TLS-N~\cite{ritzdorftls} enhances the TLS protocol
by providing non-repudiation. TLS-N authenticates TLS records sent to clients
during client-server TLS sessions. TLS-N requires TLS stack modifications and
provides hard-to-process data feeds, but it does not introduce any new trusted
entities.

In this paper, we propose \name, a practical data feed service for smart
contracts that aims to fill the gap between oracle solutions and transport-layer
authentication. Our architecture allows content providers to link their web
entities with their blockchain entities.  This design provides many benefits
like security, efficiency, and possible new features.  In \name, data is
authenticated over blockchain but without breaking TLS trust chains or modifying
TLS stacks. Moreover, content providers can specify data formats they would like
to use freely; thus data can be easily-parsable and tailored for smart
contracts. Besides that, \name provides content providers with a payment
framework, but it does not allow content providers to misbehave by equivocating
or censoring queries.

%% file: sec/pre.tex
\subsection{Blockchain and Smart Contracts}
\label{sec:pre:blockchain}
Bitcoin~\cite{nakamoto2008bitcoin} introduced the concept of open and 
decentralized consensus which, in combination with an append-only data 
structure, leaded to the existence of cryptocurrency without trusted parties. 
This combination and its variants are usually referred to as a blockchain. 
Bitcoin has inspired other systems (e.g., Litecoin~\cite{litecoin} and Namecoin 
\cite{namecoin}). Interesting and promising platforms leverage blockchain to 
implement smart contracts. These systems rely on the append-only property 
provided by blockchain platforms that allow realizing smart contracts by a 
replicated execution (i.e., all participants execute the same code for the same 
inputs, thus maintaining the same state). Those platforms introduce high-level
languages that allow to specify agreements by any parties and execute these
agreements on top of the blockchain.

The most prominent smart contract platform is Ethereum~\cite{wood2014ethereum}.
It follows the replicated execution model, and it provides smart contract 
oriented high-level languages. In Ethereum, anyone can specify a smart contract 
(i.e., an object with a set of methods and an associated state) and deploy 
it on the blockchain (each smart contract gets a unique blockchain address). 
From this point, anyone can interact with the contract by sending transactions 
to its address and calling its method(s). Smart contracts can implement almost 
arbitrary logic, including monetary transfers, thus making this technology 
appealing to financial related services and other businesses.

\subsection{Transport Layer Security}
The Transport Layer Security (TLS) protocol~\cite{dierks2008transport} is one of
the most widely deployed security protocols on the Internet. The protocol is
designed for the client-server architecture. TLS aims to provide data
confidentiality and integrity and authentication of protocol participants, 
but it was not designed to provide non-repudiation. Therefore, a communicating 
party (i.e., a client or a server) cannot prove to any third party that a given 
content was produced during the TLS connection. The TLS is prominently deployed
for securing web traffic (i.e., HTTPS).

Authentication in TLS is based on the X.509 public-key infrastructure
(PKI)~\cite{RFC5280}. Every entity that wishes to get its identity
authenticated has to obtain a digital certificate asserting the identity and
its public key. Certificates are issued by trusted entities called
certification authorities, which are obligated to verify the identity of a
requester and issue a certificate correspondingly. During a TLS connection
establishment, a server presents its certificate to the client which verifies 
the certificate and the server's identity and then uses the corresponding 
public key to continue an agreement of a shared secret key. This key is used
for protecting the subsequent communication.

\subsection{Tamper-evident Data Structure}
\label{sec:pre:logging}
A Tamper-evident Data Structure (henceforth as \teds) is a data structure that 
allows building log systems where an untrusted logger records clients' entries in 
an append-only log. The logger must be able to prove to auditors that:
\begin{enumerate*}
    \item every logged entry is still present in the log, and
    \item one snapshot of the log is consistent with any its previous version.
\end{enumerate*}

Many early proposals aimed to achieve similar properties, mainly in the context
of building a digital notary~\cite{haber1990time,bayer1993improving,goodrich2012efficient}. 
However, the semantics of \teds and multiple efficient constructions to achieve 
it were proposed by Crosby and Wallach~\cite{crosby2009efficient}. In their 
system \teds is based on a Merkle tree~\cite{merkle1989certified} (also called 
a hash tree). A Merkle tree is a binary tree where leaf nodes are labeled with 
the hash of entries and non-leaf nodes are labeled with the hash of the 
concatenated labels of its child nodes. Therefore, the \textit{root} of the 
tree is an aggregated integrity information about all its leaves.

In the Crosby-Wallach construction, the log structure is a Merkle hash tree 
with submitted entries as the leaves. The log is append-only, i.e., the entries 
are sorted in chronological order of their submission, and no leaf can be 
retrospectively removed or modified. The log supports the following
history-related operations (we give examples of these operations in
\autoref{sec:details:content_contract:update}):

\textbf{Addition} of an entry. Whenever a new entry is added to a log, a new
leaf is added to the tree, and the tree is re-computed (entries can be added in 
batches, so that the tree need not re-compute for every single entry). Adding 
new data entries requires re-computing O($log\ n$) nodes, where $n$ is the 
number of log entries.

\textbf{Membership Proof Generation} for an entry produces a
\textit{membership proof} that proves that it is part of the log. The 
membership proof of an entry is the minimal set of tree nodes (i.e., 
hashes) required to reconstruct the root. In the described construction, a 
membership proof requires O($log\ n$) nodes.

\textbf{Membership Verification} for a given entry verifies whether the entry
is part of the given log snapshot. It takes an entry, a membership proof, and a
root value as input and verifies whether the entry matches the proof and 
whether the proof terminates at the given root (i.e., the computed path has the 
root at the end). The operation returns True if the verification is successful 
and False otherwise. It is efficient since it only requires O($log\ n$) hash
operations.

\textbf{Consistency Proof Generation} for two different snapshots of the log,
a newer and an older, provides a short proof (i.e., O($log\ n$) nodes)
that the newer snapshot is an extension of the older one, i.e., the
newer snapshot was produced by only appending entries to the older
snapshot.

\textbf{Consistency Verification} takes as an input a consistency proof
between two snapshots and verifies whether the consistency proof is
correct, i.e., whether indeed the new version of the log was obtained by
appending new entries. The verification procedure is also efficient
(i.e., logarithmic in time and space) with respect to the log's size.

%% file: sec/overview.tex
\subsection{System Model}
There are the following parties in a \name system:

\textbf{Content Providers} are entities that provide content. For a simple and
intuitive description, we assume that the content is provided through the 
secure web (HTTPS); however, such a setting is not mandatory, and content 
providers do not have to run web services. Domain names identify content 
providers, and their content is accessed through URL addresses. Each content 
provider has a valid TLS certificate. In essence, content providers are not 
different from today's websites.

\textbf{Contract Parties} are mutually untrusting parties that would like to
arrange a smart-contract-based agreement which requires data from a content 
provider. Contract parties have to agree on who can act as the content provider 
for their \textit{relying contract}. Therefore, content providers are trusted 
only locally by parties that want to trust them. We assume that the protocol 
parties have access to a blockchain platform with smart contracts enabled 
(e.g., Ethereum).

We assume an adversary whose goal is to produce fake data on behalf of a content
provider. The adversary can eavesdrop, modify, and inject any protocol messages.
She can also interact freely with protocol parties and the blockchain platform.
We assume that the adversary cannot compromise underlying cryptographic
primitives and protocols (i.e., TLS), and cannot violate properties of the deployed
blockchain platform. Moreover, we assume that the adversary cannot compromise
content providers' secret keys (i.e., the one used to interact with the 
blockchain, also known as wallet private key) and cannot obtain a malicious 
certificate for a content provider (i.e., cannot compromise the TLS PKI). 
However, we discuss such strong adversaries in \autoref{sec:analysis}.

We also assume a content provider trying to misbehave by launching an
equivocation attack~\cite{tomescu2017catena} or by censoring queries for its
content. In the former case, the content provider should not be able to modify
or delete any published content retrospectively. For the latter case, censorship
is especially important in the context of the smart contract, as a content
provider could influence a contract execution by censoring some required
content. Thus for this attack, censorship attempts should be at least visible.

\subsection{Desired Properties and Design Space}
\label{sec:pre:properties}
Below we list the desired properties of a data feeds service for smart
contracts.

\textbf{Easily parsable data feeds:} data feeds should be easily parsable by
    smart contracts which use them. Besides practical implications like a
    more straightforward code base, this property improves the 
    cost-effectiveness of smart contracts deployment, as smart contract 
    platforms usually \textit{charge} contract executions per number of 
    operations.
    
\textbf{Authenticity of data feeds:} the high evidence that data feeds are
    authentic (i.e., were produced by a content provider trusted by contract
    parties) should be provided. Ideally, authenticity verification should
    follow a direct and natural trust chain (i.e., contract parties trusting
    \url{example.com} can specify in their contract that the contract can
    rely only on data provided by \url{example.com}).
    
\textbf{Easy to adopt and deploy:} all protocol parties (including content
    providers) should be able to start using the data feed system without
    major changes like requiring new infrastructure or non-backward
    compatible changes to lower-layer protocols. Ideally, the system should
    be implementable and deployable in today's setting with existing
    protocols and infrastructures.

\textbf{Non-equivocation:} Data feeds should be unable to modify or delete
    content retrospectively once data are committed and published. It enforces 
    a content provider to verify and guarantees the correctness of data before 
    performing publications. Preferably, providers should implement data 
    structures that are append-only for their publications database. 

\subsection{High-level Overview}
Design decisions behind \name try to achieve all stated properties above. First of all, 
in our system non-repudiation is provided directly by content providers. This is 
similar to the approaches that modify the TLS protocol; however, the authentication
is not conducted at the TLS layer. Instead, we introduce a layer of indirection 
that allows authenticating content on the blockchain.

\begin{wrapfigure}{R}{0.45\textwidth}
    \vspace*{-0.6cm}
    \centering
    \includegraphics[width=\linewidth]{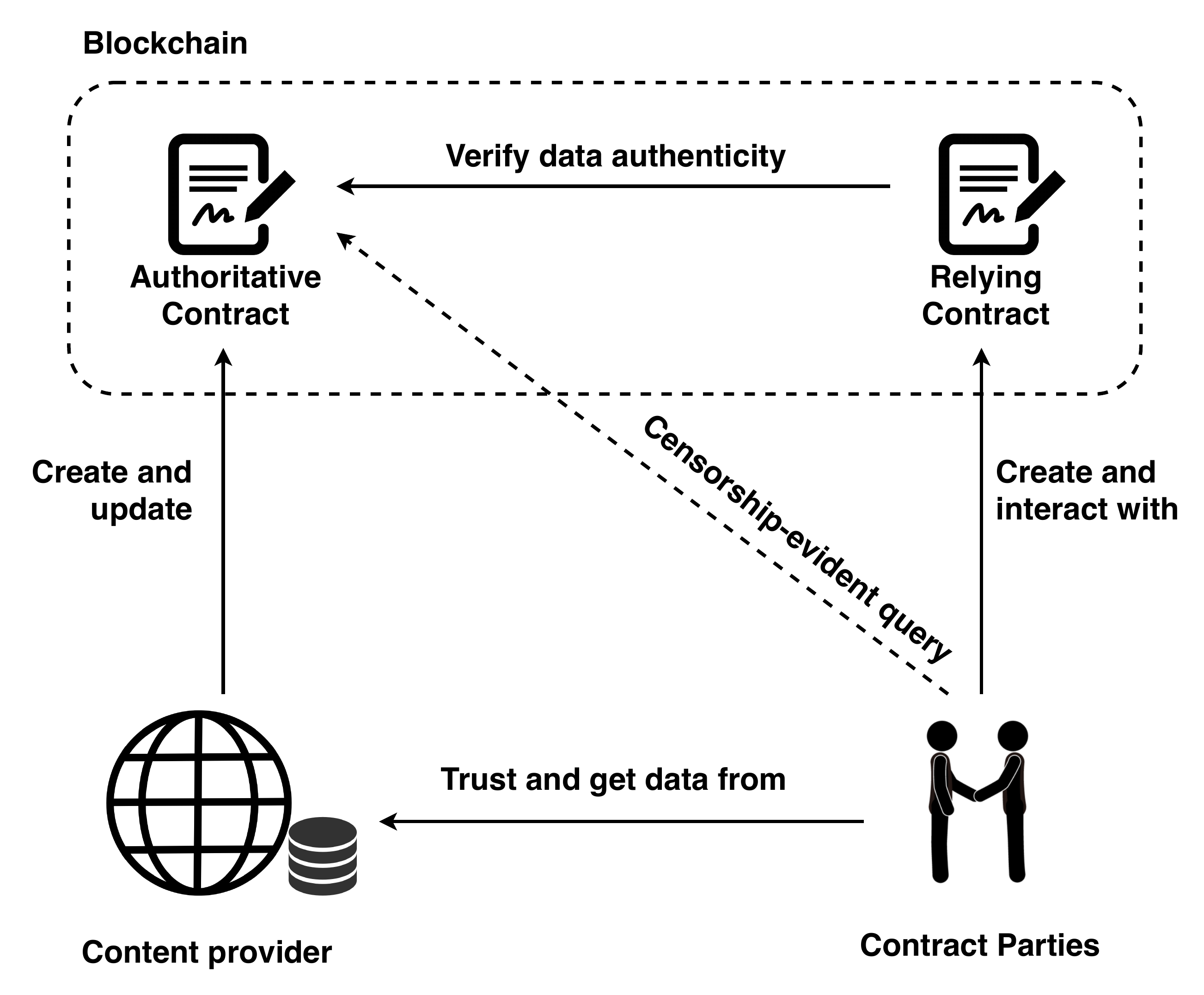}
    \caption{High-level overview of \name.}
    \label{fig:sysmodel}
    \vspace*{-0.6cm}
\end{wrapfigure}

In our design, content providers link their TLS identities with their blockchain
identities and the locations of special smart contracts used for authenticating and
verifying their content. Such a design provides multiple benefits. Firstly, it
enables verifying blockchain identities, directly through the existing TLS PKI.
Secondly, it allows relying contracts to validate the authenticity of data as
simple as calling another smart contract's method (without involving any
in-contract expensive public-key operations). Lastly, integrating content
providers with blockchain enables new features like keeping the providers
accountable, proving their unavailability or providing a payment framework that
can incentivize them to initiate the service. A high-level overview of our 
system is shown in \autoref{fig:sysmodel}, and in this section, we describe its 
steps and the main components.

The first step in our protocol is to create a \textit{authoritative contract} by a
content provider who wishes to participate in \name. The main aim of authoritative
contracts is to enable other contracts to verify the authenticity of the content
produced by content providers. Authoritative contracts provide additional
functionalities by ensuring that content providers do not misbehave: a) by
retrospectively tampering with their data, or b) by censoring queries sent to
them.

Every authoritative contract provides an API that allows:
\begin{enumerate*}
    \item its owner (i.e., the content provider) to update it,
    \item other contracts to verify that the content provider indeed produced
    given data,
    \item contract parties to make \textit{censorship-evident queries} to the
    content provider for the specific content (this option is used when the
    content provider seems unavailable or is censoring some queries).
\end{enumerate*}

In the second step, the content providers create a signed \textit{manifest} that
contains the following elements:
\begin{enumerate*}
    \item a location (i.e., a blockchain address) and interface structure of its
    authoritative contract,
    \item metadata specifying details of provided content.
\end{enumerate*}
The manifest is signed, and the manifest's signature is computed using the private
key corresponding to the public key from the content provider's TLS certificate.
Such a setting follows the natural trust chain; therefore, it allows contract
parties to verify the authenticity of manifests directly, using the TLS PKI, and
without breaking existing trust chains.

The content provider creates a \teds that will store data entries that the
content provider wants to serve. The first entry of this data structure is the
manifest. Although \name data may be published using HTTPS services, those
services focus on data privacy and integrity. We define that the manifest must 
be signed and added into the \teds to extend security properties including
non-repudiation and non-equivocation to it.

For every update, the content provider adds new data entries to its
\teds, re-computes the data structure, and sends the new root and its
corresponding consistency proof to the authoritative contract (they do
not store any actual content, but only \teds roots --- the short authentication
information about the content.) The authoritative contract validates the sent 
information enforcing the append-only property (i.e., it makes sure that the 
content provider is appending data only -- not modifying nor removing any 
entries).  The data entries with their corresponding membership proofs are 
published at a pre-defined URL location, so that everyone can locate and access it.

Contract parties that would like to deploy a \textit{relying contract} (i.e., a
smart contract which depends on a data feed from an external website) have to
find and agree on a content provider (this process is realized out of band).
When contract parties find the content provider they would like to use, they 
locate and verify its manifest and authoritative contract, and associate the 
location of the authoritative contract as an oracle in their relying contract.

Whenever one contract party would like to call a method that uses content
provider's data, it accesses the required data entry and its membership proof
from the content provider and then calls this method with this pair (and a fee
for content provider) as the arguments. Now, the method needs to verify whether
the content provider indeed produced the data entry and to do so, the relying
contract only requires to call the authoritative contract's membership verification
method. When the data entry is verified, the relying contract's method can
continue with its processing logic.

%% file: sec/details.tex
\begin{wrapfigure}{R}{0.60\textwidth}
    \vspace{-0.6cm}
    \centering
	\includegraphics[width=\linewidth]{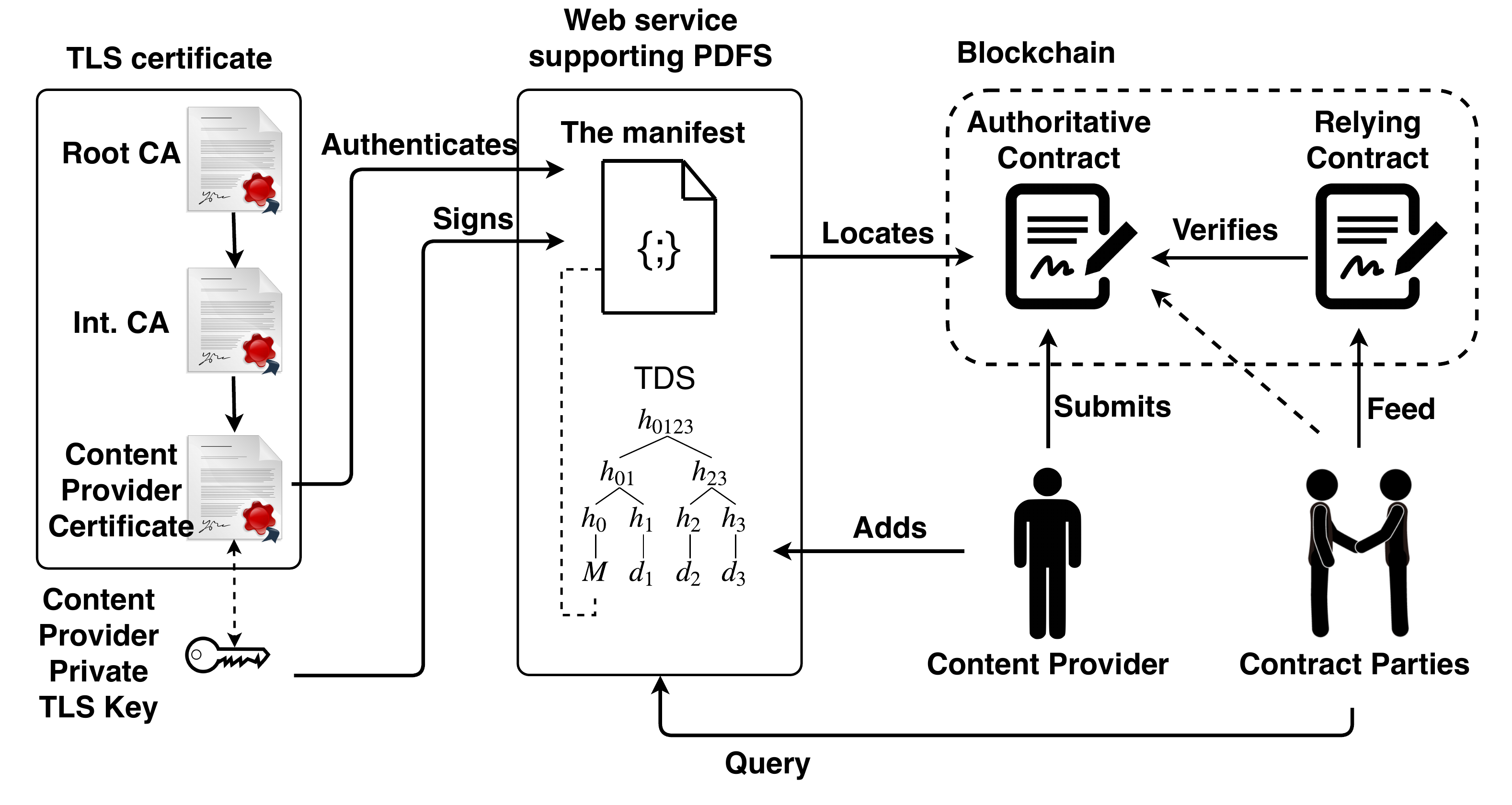}
	\caption{Details of the \name architecture and parties interactions.}
    \label{fig:pdfs_interactions}
    \vspace{-0.6cm}
\end{wrapfigure}

In this section, we describe components of the \name architecture and explain
its different steps from a content provider establishing its \name service
until contract parties using the provider's data to make a transaction within
their smart contract. We also discuss how the content provider maintains the
service. As shown in~\autoref{fig:pdfs_interactions}, a \name service consists
of an authoritative contract, a web service whose entries are kept within a \teds, and
a manifest. We provide details of these components and their functionality in
this section.

\subsection{Service Initialization}
In the first step, the content provider initializes a \name service by 
deploying an authoritative contract in the blockchain. This contract is 
designed to interact with the content provider's back-end service, relying 
contracts, and contract parties. Initially, the authoritative contract has 
empty storage; however, it will store root hashes of the deployed \teds. These 
root hashes will enable the contract to check on demand the consistency 
between two \teds snapshots (i.e., ensuring that the content provider updates 
its \teds correctly) and to conduct a membership verification  (i.e., verifying 
for relying parties that an entry is part of the content provider's \teds). 
Further details of authoritative contracts are discussed in
\autoref{sec:details:content_contract}. Once it is deployed, the content 
provider gets an address of the authoritative contract instance.

Then, the content provider creates a manifest. The manifest is a file that
describes details of the \name service. It is necessary for contract parties,
since based on the manifest, they can create a workable relying contract. The
manifest has to be authentic. Therefore, the content provider signs it. As TLS
certificates issued by CA are widely trusted parties on the Internet, the 
content provider can sign the manifest using the private key corresponding to 
its TLS certificate for supporting HTTPS web traffic. Such a design choice has 
multiple benefits. Firstly, it simplifies the signature creation and 
verification process since contract parties can obtain the required certificate 
by visiting the content provider's website. Secondly, the manifest is 
authenticated following an already existing trust chain. When the manifest is 
signed, it is added as the first element to the content provider's
\teds. We define and describe the fields that a manifest contains:

\textbf{URL} corresponds to the URL address used by the content provider to
    publish data, and it indicates where contract parties can access data
    entries.

\textbf{Authoritative Contract Address} is the address in the blockchain 
    associated with the deployed authoritative contract. Contract parties 
    preload their relying contract with the value of this field (to allow them 
    calling procedures or functions on the authoritative contract instance).

\textbf{Authoritative Contract Interface} is an abstract structural descriptor 
    of the authoritative contract. It includes definitions of functions, access 
    method, and parameters. Likewise the authoritative contract address, data 
    contained in this field has to be embedded in the relying contracts as an 
    object interface. This field is platform dependent (e.g., the ABI in 
    Ethereum).

\textbf{Data Structure} describes the encoding or structure of data entries
    that the content provider stores in its \teds. Typically, content providers
    use widely adopted data encodings, such as JSON or XML. Thus, the content 
    provider presents here which values and data types are expected to be found
    within every data entry. This field is necessary for contract parties to 
    understand the semantics of data entries and to create their relying 
    contracts able to parse data entries and implement their processing logic 
    correctly.

\textbf{Signature} is a field that authenticates all values contained in the
    manifest. As described above, the signature is computed using the private
    key associated with the content provider's TLS certificate.

If the TLS certificate expires, the \name service is not affected for relying 
contracts already deployed. It is because contract parties use the certificate
to verify the manifest signature before they create relying contracts. 
Furthermore, neither the authoritative contract nor relying contracts perform 
any signature verification later. Also, the content provider does not require to 
terminate the \name service if the TLS certificate is reissued using the same 
private-public key pair that was used in the manifest creation.

\subsection{Authoritative Contract}
\label{sec:details:content_contract}
The authoritative contract is a central point in the \name architecture. It 
interacts with the content provider back-end, relying contracts, and contract 
parties. Its primary goal is to ensure that the content provider indeed 
published a specific data entry. A detailed pseudo-code of the authoritative 
contract is shown in Alg.~\autoref{alg:content:contract}. An authoritative 
contract consists of the functions that allow:

\begin{compactitem}
  \item The content provider to store root hashes once the consistency is 
      verified. This procedure is executed by calling the \textsc{update} 
      function (details about the consistency verification in
      \autoref{sec:details:content_contract:update}). The \textsc{update}
      function can be executed only by the content provider. For efficient 
      storage management and time delays or race conditions avoidance, the 
      authoritative contract only stores an array of the last $K$ root hash 
      values committed ($K$ is defined by the content provider).
  \item Relying contracts to make trustworthy transactions based on data entries
      whose origin and integrity are verified by calling the \textsc{membership}
      function. This function checks whether a data entry and its membership
      proof is valid comparing to stored roots.
  \item Contract parties to make censorship-evident queries using the 
      \textsc{query} function and get responses by calling the 
      \textsc{get\_response} function. These queries and responses are sent over 
      the blockchain, therefore they are publicly visible.
\end{compactitem}

\begin{algorithm}[t]
    \scriptsize
    \begin{multicols}{2}
    \begin{algorithmic}[1]
    \Statex
    \begin{compactdesc}
    \item[$FEE_{mem}$:] the cost for membership verification,
    \item[$FEE_{query}$:] the cost for making a censorship-evident
    query,
    \item[locked:] boolean value that indicates whether the authoritative contract
    can be updated,
    \item[roots:] a map of roots hashes; it uses a timestamp as the key,
    \item[time:] a value that indicates the last updating time,
    \item[queries:] a map of censorship-evidence query made; it uses a
    number as the key,
    \item[responses:] a map of responses for queries made; the key is
    associated to existing identifiers in the queries map,
    \item[counter:] an incremental number used as the identifier for
    the queries made,
    \item[\textsc{NOW}():] the current block timestamp,
    \item[\textsc{HASH}():] a cryptographic hash function.
    \end{compactdesc}
    \Statex
    \Procedure{init}{}
        \State $roots \gets \emptyset$
        \State $time \gets 0$
      \State $locked \gets False$
    \EndProcedure
    \Procedure{update}{$root,proof_{cons}$}
        \Assert{$sender = owner$}
      \Assert{$locked = False$}
        \If {$\textsc{CONSISTENCY}(root,proof_{cons})$}
          \State $time \gets \textsc{NOW}()$
            \State $roots[time] \gets root$
        \EndIf
    \EndProcedure
    \Procedure{lock}{}
        \Assert{$sender = owner$}
        \State $locked \gets True$
    \EndProcedure
    \Procedure{consistency}{$root,proof_{cons}$}
        \If {$time = \textit{0}$}
            \State \Return true
        \EndIf
        \State $(root_{new}, root_{old}) \gets \textsc{MTH}(proof_{cons},\emptyset)$
        \State \Return $(root_{new} = root \quad \& \quad root_{old} = roots[time])$
    \EndProcedure
    \Procedure{membership}{$data,proof_{mem},fee$}
        \Assert{$fee = FEE_{mem}$}
        \State $leaf \gets \textsc{HASH}(data)$
        \State $(root_{mem}, \_) \gets \textsc{MTH}(proof_{mem},leaf)$
        \State \Return $root_{mem} \in roots$
    \EndProcedure
    \Procedure{mth}{$proof,leaf$}
        \State $i \gets 0$
        \State $hash_{x} \gets hash_{y} \gets leaf$
        \If {$leaf = \emptyset$}
            \State $i \gets 1$
            \State $hash_{x} \gets hash_{y} \gets proof_{(0)}.{hash}$
        \EndIf
        \For {$i < \textsc{len}(proof)$}
            \If {$proof_{(i)}.{side} = \textit{RIGHT}$}
                \State $hash_{x} \gets \textsc{HASH}(hash_{x} || proof_{(i)}.hash)$
            \Else
                \State $hash_{x} \gets \textsc{HASH}(proof_{(i)}.hash || hash_{x})$
                \State $hash_{y} \gets \textsc{HASH}(proof_{(i)}.hash || hash_{y})$
            \EndIf
          \State $i \gets i + 1$
        \EndFor
        \State \Return $(hash_{x}, hash_{y})$
    \EndProcedure
    \Statex
    \State // Censorship Evidence functions
    \Procedure{query}{$filter,fee$}
        \Assert{$fee = FEE_{query}$}
        \State $counter \gets counter + 1$
        \State $queries[counter] \gets filter$
        \State \Return $counter$
    \EndProcedure
    \Procedure{store\_response}{$id,data$}
        \Assert{$sender = owner$}
        \Assert{$id \leq counter$}
        \State $responses[id] \gets data$
    \EndProcedure
    \Procedure{get\_response}{$id$}
        \Assert{$id \leq counter$}
        \State \Return $responses[id]$
    \EndProcedure
    \end{algorithmic}
    \end{multicols}
    \caption{Authoritative Contract Pseudo-Code.}
    \label{alg:content:contract}
\end{algorithm}

\begin{figure*}[h!]
    \centering
	\includegraphics[width=\linewidth]{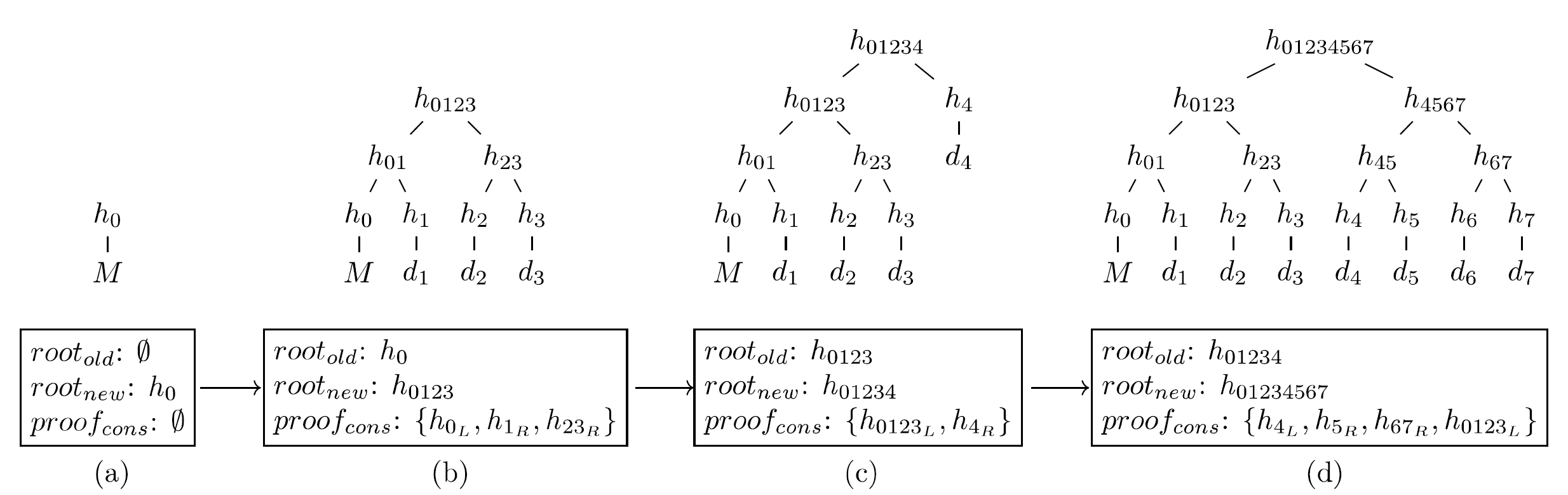}
    \caption{An example of maintaining a \teds. It is a representation of
    information provided for the consistency verification when a new snapshot of
    the \teds is updated to the authoritative contract. Each element of the
    $proof_{cons}$ indicates the hash value and the corresponding side
    ($h_{x_{L}}$ refers left position and $h_{x_{R}}$ refers right position).}
    \label{fig:tree_maintain}
\end{figure*}

Functionalities offered to contract parties are designed to require payments 
for their executions. It allows content providers to adopt a new business model
receiving payments for providing data over a \name service.\footnote{Fees for 
executing \name functions are different from fees for executing transactions on 
the blockchain (e.g., Ethereum gas cost).}

\subsection{Data Update}
\label{sec:details:content_contract:update}
Adding new data entries to the \teds requires re-computing the root. To run 
\name service properly, it also requires synchronization of changes between 
the content provider back-end (maintaining the \teds) and the authoritative 
contract which has to be updated to enable the membership verification of any 
newly added entry. To synchronize, the content provider submits the new root 
hash value along with a corresponding proof for the consistency verification. 
This verification uses the provided proof to re-calculate two hash values. and 
then, it compares those calculated hashes checking whether they are equal to 
the new root value to store and the last one stored in the authoritative 
contract accordingly. This guarantees that the new \teds is an extension of the 
last one committed confirming that no previous data entry has been altered or 
removed. If there is an error, the authoritative contract ignores the submitted 
data and remains in the current state. Once the new root is accepted by the 
authoritative contract, the content provider can make the updated \teds 
accessible over HTTPS.

In~\autoref{fig:tree_maintain}, we show an example of how a \teds evolves when 
data entries are added, and what values are sent for submitting roots to the
authoritative contract. In case $(a)$, the new root is directly stored with no 
previous validation as it is the first one, and there is no consistency to 
evaluate. In case $(b)$, the new root is submitted along with the following 
consistency proof ($proof_{cons}$). The authoritative contract uses the 
provided data to evaluate the \teds consistency. In this case, the consistency 
verification is easy to deduce since the previous root ($h_{0}$) is contained 
in the provided proof. Similarly in the case $(c)$, the previous root 
($h_{123}$) is contained in the $proof_{cons}$ array.

However, the case $(c)$ shows a particular situation due to the \teds is 
unbalanced. It changes how the consistency verification works for the next 
root submission, the case $(d)$. For it, the consistency proof provided is: 
$proof_{cons} = \{h_{4_{L}}, h_{5_{R}}, h_{67_{R}}, h_{123_{L}}\}$. Because of
the unbalanced \teds, the consistency verification re-calculates both roots, 
the previous one ($h_{1234}$) and the new one ($h_{1234567}$) by using the 
same provided proof. To calculate the previous root, the consistency 
verification only needs the contained elements $\{h_{4_{L}}, h_{123_{L}}\}$. 
Furthermore, the complete array is used to re-calculate the new root. 
Therefore, the procedure can confirm the consistency of the new \teds.

\subsection{Relying Contracts}
\label{sec:details:relying_contract}

A relying contract is a smart contract which is created by contract parties and
needs content providers data to validate conditions and perform transactions.
Before it is created, contracts parties agree on a content provider they trust
which provides a \name service. After validating its manifest signature, 
contract parties extract the information contained in the manifest and use it
to prepare and deploy a relying contract. In that way, the relying contract 
will interact with the correct authoritative contract and be able to: 
\begin{enumerate*}
    \item execute the membership verification procedure,
    \item get the response for a censorship-evident query, and
    \item parse data entries and execute a processing logic depending on data 
    entry fields.
\end{enumerate*}
We provide a pseudo-code example of a relying contract in Alg.
\autoref{alg:relying:contract}.

When needed, contract parties request a specific data entry to the content
provider, which responses a data entry along with its respective membership
proof. Considering case $(c)$ in~\autoref{fig:tree_maintain}, let us assume the
content provider is queried for the data entry $d_2$, so its response will 
contain the asked data entry $d_2$ along with a membership proof $proof_{mem} = 
\{h_{3_{R}}, h_{01_{L}}, h_{4_{R}}\}$. Once that data is submitted to the 
relying contract, it will execute the membership verification intreating with
the authoritative contract. As we see in this example, the provided proof and 
the data entry's hash value lead to re-calculate the root $h_{1234}$ which is 
stored in the authoritative contract and it confirms data authenticity. If any 
value is modified, either the data or the proof, the membership verification 
re-calculates a different hash value which does not correspond to any stored 
root, so the verification fails.

\vspace{-0.4cm}
\begin{algorithm}[h!]
    \scriptsize
    \begin{multicols}{2}
    \begin{algorithmic}[1]
    \Statex
    \begin{compactdesc}
    \item[$cc$:] authoritative contract object interface.
    \end{compactdesc}
    \Statex
    \Procedure{init}{$addr$}
        \State $cc\gets \textsc{Authoritative\_Contract}(addr)$
    \EndProcedure
    \Procedure{submit\_data}{$data,proof_{mem},fee_{mem}$}
        \State $v \gets False$
        \State $v \gets cc.\textsc{membership}(data,proof_{mem},fee_{mem})$
        \If {$v = True$}
        \State $\dots$ Decode data input
        \State $\dots$ Decide and make transaction
        \EndIf
    \EndProcedure
    \Procedure{if\_censorship}{$id$}
        \State $data \gets cc.\textsc{get\_response}(id)$
        \If {$data \not= \emptyset$}
        \State $\dots$ Decode data input
        \State $\dots$  Decide and make transaction
        \EndIf
    \EndProcedure
    \Statex
    \State \textbf{interface} \textsc{Authoritative\_Contract:}
        \State $\quad$ \textbf{procedure} $\textsc{membership}(data,proof,fee)$
        \State $\quad$ \textbf{procedure} $\textsc{get\_response}(id)$
        \State $\quad$ $\dots$ Any additional procedure defined
    \end{algorithmic}
    \end{multicols}
    \caption{Relying Contract Template.}
    \label{alg:relying:contract}
\end{algorithm}
\vspace{-0.6cm}

\subsection{Censorship Evidence}
\label{sec:details:censorship}
Censorship is an especially challenging threat since a content provider 
censoring queries can influence executions of agreements based on smart 
contracts, and censorship is difficult to prove. However, \name extends 
the authoritative and the relying contract with functions to allow 
\textit{censorship-evident queries}. So contract parties can query a 
content provider over the blockchain whenever they cannot obtain data 
directly through conventional channels (e.g., like HTTPS). All interactions, 
contract parties' query and content provider's response, are recorded as 
transactions in the blockchain. Therefore, they are visible for anyone, and any
censorship attempt is publicly observable. We discuss censorship attacks 
further in~\autoref{sec:analysis:malicious}.

\subsection{\name Service Termination}
\label{sec:details:termination}
Content providers might need to terminate a \name service due to operational
management or security reasons. To do so, they can execute the \textsc{lock} 
function which disallows any future update attempt of the authoritative 
contract. Locking authoritative contracts does not introduce collateral damage 
to already-deployed relying contracts. A locked authoritative contract can be 
used for membership verifications as long as the corresponding root value is 
stored. In particular, the locking function might be useful in the case of a 
security breach (like a stolen blockchain private key), to prevent an adversary 
from submitting malicious root values (we discuss details in 
\autoref{sec:analysis:stolen}).

%% file: sec/analysis.tex
In this section, we discuss different attacks and their implications over 
\name. However, this discussion is extended in
\autoref{sec:appendix:security} in the appendix which also addresses issues and 
disagreements that one might argue against our proposed solution.

\subsection{PKI and Key Compromise}
\label{sec:analysis:stolen}
An adversary able to compromise the TLS PKI can create a malicious manifest and
an authoritative contract, and can impersonate the content provider by creating
arbitrary content. Interestingly, even if successful, such an adversary cannot
undermine the security of the relying contracts already deployed since these
contracts use the \textit{correct} authoritative contract instance for data verification.
Moreover, by deploying a new (malicious) authoritative contract, the adversary needs
to deploy it over the blockchain, which makes the attack visible and detectable.

A more severe attack is a compromise of the private key used for the
interactions between the content provider and the blockchain platform. In such
a case, the adversary can add to the existing \teds malicious entries, re-compute 
the structure, and update the authoritative contract with a new root. Then, these 
malicious entries can be used by relying smart contracts for processing. However, 
even in that case the attack is visible since the authoritative contract is 
updated publicly, on the blockchain. Thus, the content provider will notice it 
and terminate its service (see~\autoref{sec:details:termination}).

\subsection{Malicious Content Provider}
\label{sec:analysis:malicious}

\name prevents and mitigates some attacks conducted by a malicious content 
provider. The design of authoritative contracts in \name does not allow the 
content provider (or an adversary with the content provider's blockchain key) 
to retrospectively modify or remove content. The authoritative contract 
enforces the consistency of the \teds for every update (see 
\autoref{fig:tree_maintain}). This property is also crucial for thwarting 
equivocation attacks~\cite{tomescu2017catena}. A manifest file identifies the 
authoritative contract that guarantees that the content provider cannot 
equivocate as long as the blockchain platform is secure (see 
\autoref{sec:analysis:blockchain} in the appendix). The content provider can 
create multiple manifest files and authoritative contract, however, a) it does 
not influence already deployed contracts, b) is not necessarily a malicious 
activity, and c) is visible over the blockchain; thus, it can be monitored.

\name provides non-equivocation by ensuring that content providers'
database is append-only. However, it does not prevent a content provider from
adding two semantically conflicting entries to their databases (e.g., two
different results for a same football game). Conflicting entries can be harmful 
to relying contracts as they may lead to completely different execution paths. 
Since \name does not allow content providers to ``overwrite'' their entries, we 
suggest that such conflicts should be handled by relying contracts themselves. 
More precisely, using agreement protocols like implementing \textit{grace 
periods} or submitting data from \textit{multiple content providers} before 
making final decisions, such that any conflicting entry submitted can reverse 
contracts agreements.

A subtler attack is a content provider censoring queries. That risk is
especially important, when a malicious content provider ignores contract 
parties' queries, pretending unavailability or displaying incorrect data that 
cannot be successfully verified by relying contracts. In such a case, \name 
allows contract parties to query the content provider over the blockchain for a 
required query (see~\autoref{sec:details:censorship}). The content provider is 
obligated to response due to the query and content provider's response are 
publicly visible.

%% file: sec/implementation.tex
In this section, we demonstrate that \name fulfills the desired properties 
explained in~\autoref{sec:pre:properties}. We fully implemented a proof of 
concept which involved both parties of a \name architecture (the content 
provider and contract parties). Although we tested \name under a generic 
scenario (see~\autoref{sec:appendix:casestudy} in the appendix), \name can be 
integrated into any context where smart contracts need to make decisions
based on external data. Our solution allows content providers, regardless of 
the content and data type, to become a trustworthy data feed for smart 
contracts.

\subsection{Implementation}
To approach our implementation of \name, we developed a web service for the
content provider using Go v1.10.1 as the programming language. It is a RESTFul
API which offers data entries encoded in JSON format. This application is 
configured to support HTTPS, and we deployed a private PKI infrastructure and 
TLS certificates using OpenSSL v1.1. For contract parties, we implemented a
client in Python v3.6.5 which is able to request data entries to the created 
web service. Smart contracts, the authoritative and the relying contract are
coded in Solidity v0.4.21 and deployed in an Ethereum blockchain. To allow 
reproducibility of our experiments and evaluations, we publish our 
implementation at~\url{https://gitlab.com/juan794/pdfs}.

\subsection{Evaluation}
\label{sec:implementation:evaluation}
In this section, we discuss results obtained from a series of experiments we
performed. To evaluate \name, we used a computer which has 16GB of RAM and a 
CPU Intel Core i7 7700H. We performed measurements regarding the execution 
cost which is expressed in Ethereum gas units, and then, converted to US 
dollars.

\begin{figure}[t!]
    \centering
    \includegraphics[width=\linewidth]{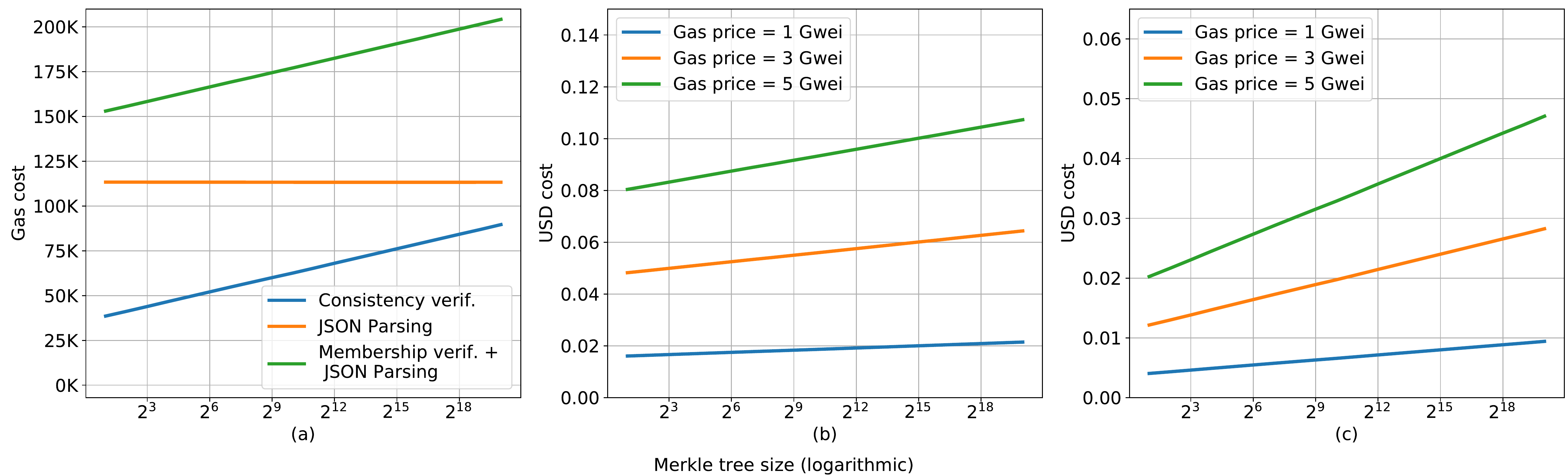}
    \caption{Ethereum gas consumption and price variation analysis converted to US 
    dollars. (a) Gas cost of \name operations (b) membership verification cost (c) 
    consistency verification cost.}
    \label{fig:analysis}
\end{figure}

We analyzed the cost growth according to the number of data entries in the
\teds. As shown in~\autoref{fig:analysis}, we observe that the cost for the
consistency and membership verification grows on a logarithmic scale as expected
since we deployed a \teds using binary Merkle trees. In the case of the JSON
parsing, the cost is constant and does not change with the \teds size.
We also disaggregate total costs to investigate the details for executing \name
procedures (see details in~\autoref{tab:benchmark}). In the case of 
having a data feed with more than 1 million ($2^{20}$) data entries, we observe 
that the consistency verification has a gas cost of 86,642 on average, where only 
4\% of this cost is related to the hash calculations. The remaining percentage
corresponds to miscellaneous code, including storage and control statements,
such as \emph{asserts}. Moreover, we also measured the cost of executing a
membership verification, and we observe that it has an average gas cost of
204,242. However, as JSON parsing is not natively supported in Ethereum, 55\% of
the total cost is spent on performing this task. On the other hand, the \emph{gas}
consumptions are 813,111 and 4,355,638 respectively for the authoritative contract 
and the relying contract deployment.

Next, we show in~\autoref{fig:analysis} what would be the maximum cost
considering the two prices involved. For our measures, we assumed a price of 5
Gwei per gas unit and a price of US\$105.05 per ether; those
are maximum conversion rates presented at the writing time. As a result, the
consistency verification costs around US\$0.048 in a \name service that contains
more than 1 million data entries. This means a cost of US\$1.7x$10^{-7}$ per
data entry. On the other hand, the membership verification of one data entry in
a \teds of that size ($2^{20}$) costs around US\$0.11. We recall that it is
including the JSON parsing which is a costly task on smart contracts. Therefore,
we show that \name is costly viable  to create and deploy a trustworthy data
feed for a smart contract. The cost can decrease if Ethereum starts supporting
JSON parsing natively or if content providers use a more efficient data entry
encoding.

\begin{table}[t!]
    \centering
    \scriptsize
    \begin{tabular}{l|rrrrr|} \hline
    \multicolumn{1}{|l|}{\teds size}  & \multicolumn{1}{c}{$2^{1}$} & \multicolumn{1}{c}{$2^{5}$} & \multicolumn{1}{c}{$2^{10}$} & \multicolumn{1}{c}{$2^{15}$} & \multicolumn{1}{c|}{$2^{20}$} \\ \hline
      & \multicolumn{5}{|c|}{Membership verification cost} \\ \hline
      \multicolumn{1}{|l|}{JSON Parsing} & 113,349 (74\%) & 113,325 (69\%) & 113,293 (63\%) & 113,273 (59\%) & 113,298 (55\%) \\
      \multicolumn{1}{|l|}{Hash calculation} & 447 (1\%) & 1,107 (1\%) & 1,933 (2\%) & 2,757 (2\%) & 3,583 (3\%) \\
      \multicolumn{1}{|l|}{Miscellaneous} & 39,253 (25\%) & 49,369 (30\%) & 61,905 (35\%) & 74,633 (39\%) & 87,361 (42\%) \\
    \multicolumn{1}{|r|}{\textbf{Total}} & 153,049 & 163,801 & 177,131 & 190,663 & 204,242 \\ \hline
      & \multicolumn{5}{|c|}{Consistency verification cost} \\ \hline
      \multicolumn{1}{|l|}{Hash calculation} & 149 (1\%) & 809 (2\%) & 1,634 (3\%) & 2,294 (3\%) & 3,284 (4\%) \\
      \multicolumn{1}{|l|}{Miscellaneous} & 38,419 (99\%) & 48,551 (98\%) & 60,961 (97\%) & 71,158 (97\%) & 86,358 (96\%) \\
    \multicolumn{1}{|r|}{\textbf{Total}} & 38,568 &  49,360 & 62,595 & 73,452 & 89,642 \\ \hline
    \end{tabular}
    \caption{Cost analysis for membership and consistency verification
    considering multiple sizes of the \teds.}
    \label{tab:benchmark}
    \vspace{-1.1cm}
\end{table}

\begin{wraptable}{r}{0.38\textwidth}
    \vspace{-0.6cm}
    \centering
    \scriptsize
        \begin{tabular}{|cccc|} \hline
            \name & secp256r1 & RSA & ECRecover \\ \hline
            87,361 & 1,854,634 & 596,287 & 38,887 \\ \hline
        \end{tabular}
        \caption{Ethereum gas consumption of \name compared to signature verifications.}
        \label{tab:benchmark:signatures}
    \vspace{-0.6cm}
\end{wraptable}

In \autoref{tab:benchmark:signatures}, we show the \emph{gas} consumption
comparing \name against signature verification algorithms, such as 
ECRecover~\cite{ecrecover} (native in Ethereum), TLS-N implementation of 
secp256r1~\cite{tlsnutils} and RSA~\cite{rsaverifier}. We observe that the Ethereum 
native function for signature verification is cheaper than \name. On the other hand, 
\name is significantly cheaper that implementations coded on Solidity programming 
language. Although those alternatives allow contract parties to verify integrity 
and provenance, they do not provide accountability or non-equivocation properties 
from content providers.

\begin{wraptable}{r}{0.48\textwidth}
    \vspace{-0.6cm}
    \centering
    \scriptsize
        \begin{tabular}{|c|rrrrrr|} \hline
            Oper.& 50B\ & 150B & 500B & 1KB & 2KB\  & 5KB\  \\ \hline
            Query & 25,597 & 32,399 & 56,337 & 90,483 & 158,644 & 363,282 \\
            Resp. & 25,804 & 32,606 & 56,544 & 90,690 & 158,851 & 363,489\\ \hline
        \end{tabular}
        \caption{The gas cost of the query and response operations.}
        \label{tab:benchmark:censor}
    \vspace{-0.6cm}
\end{wraptable}

Lastly, we investigated the cost of censorship-evident queries and responses (see
\autoref{sec:details:censorship}). As storing data in Ethereum smart contracts
is expensive~\cite{wood2014ethereum}, we implemented this functionality without
involving smart contract storage.  Instead, queries and responses are published
as blockchain transactions (as calls to the corresponding functions), but
without storing them in authoritative contracts. That improves the cost efficiency
greatly while providing the same functionality i.e., queries and responses can
be read (as they are part of the blockchain) and responses are authentic (as
they are sent within blockchain transactions signed by content providers).
The gas cost of these operations depending on a size of a query and response are
shown in~\autoref{tab:benchmark:censor}. As presented, the cost grows linearly 
with query/response's size, but queries and responses of the same size have 
roughly the same cost.

%% file: sec/related.tex
TLSNotary~\cite{TLSnotary} is a service that introduces a third-party auditor
which attests TLS session data exchanged between a client and a server.  To
provide this functionality, the protocol requires changes to the TLS protocol
like an introduction of a dedicated client-auditor protocol. TLSNotary has many
drawbacks. For instance, it is only compatible with TLS 1.0 and 1.1, while TLS
1.2 is widely deployed and recommended as default~\cite{ssllabs}. TLSNotary is
specified with obsolete cryptography algorithms, and it supports only cipher 
suites with the RSA algorithm for a secret key establishment. As TLS records 
are being authenticated, the output obtained from TLSNotary is hard to parse 
and process by smart contracts. Although, the protocol has many disadvantages, 
it got adopted by other solutions, like Oraclize~\cite{oraclize}, which 
integrates multiple data feed systems. However, as combined with TLSNotary, it
introduced a trusted third-party that holds secret keys used for auditing TLS 
sessions.

An alternative approach proposed is to use prediction markets for providing data
feeds, such as~\cite{Augur} and~\cite{Gnosis}. In such systems, users try to predict 
real-world events by betting or voting for them. Usually, these systems are 
implemented on top of blockchain platforms, hence they could be easily integrated 
with smart contracts. Unfortunately, they have many drawbacks as in the case of 
disputes there is no responsible party (i.e., responsibility is distributed). 
Moreover, data feeds depend on human inputs which can be biased, slow, or incomplete.

Town Crier (TC)~\cite{zhang2016town} takes a different approach to instantiate
data feeds for smart contracts. TC deploys trusted computing (i.e., the Intel
SGX technology~\cite{costan2016intel}) to allow special applications to interact
with HTTPS-enabled websites. In order to provide authentic data feeds, such an
application, is executed within an SGX enclave. Thus, it is possible to conduct
a remote attestation that the correct code was executed. The application
establishes a secure TLS connection with a website and parses its content,
which then can be used as an input to smart contracts. In contrast to
TLSNotary, TC can provide easy-to-parse data and is flexible since there can
be many applications. With the assumption that the contract parties have
verified an attestation of the used enclave, TC allows relying contracts to avoid
expensive public-key verifications by making assertions between enclaves and
their blockchain identities (this is a similar concept as in \name).  However,
TC has some significant limitations. First of all, it positions Intel as a
trusted party required to execute a remote attestation.  Secondly, its security
relies on the security of the SGX framework (undermined by recent severe
attacks~\cite{vanbulck2018foreshadow}) and the security of its attestation 
infrastructure, which is especially undesired as the SGX attestation 
infrastructure is a weakest-link-security system (i.e., one leaked attestation 
private key allows an adversary to attest any application). TC has inspired 
other systems, like ChainLink~\cite{ellisdecentralized}, which aims to 
decentralize TC applications by forming a network of them (to detect and deal 
with possible inconsistencies).  Unfortunately, this design does not solve the
main drawbacks of TC.

TLS-N~\cite{ritzdorftls} is a more generic approach to provide non-repudiation
to the TLS protocol. In order to realize it, TLS-N modifies the TLS stack such
that TLS records sent by a server are authenticated (in batches). Therefore,
TLS-N clients can present received TLS-N records to third parties which can
verify it, just trusting the server (without any other third trusted parties).
The main drawbacks of TLS-N are in its deployability. It requires significant
changes to the TLS protocol and as learnt from the previous deployments the TLS
standardization and adoption processes are very slow. Because of the TLS-N's 
layer of authentication, TLS records are being authenticated which is 
inconvenient and expensive to process by smart contracts. Furthermore, the TLS 
layer is uncontrollable by web developers, and thus, most of their applications 
would need to be rewritten for TLS-N. Besides that, TLS-N relying contracts 
have to conduct an authentication verification which is a costly operation.

\begin{wraptable}{r}{0.4\textwidth}
    \vspace*{-0.6cm}
    \centering
    \scriptsize
    \begin{tabular}{l|ccc|}
    \cline{2-4}
        & \shortstack{\\ No third \\ trusted \\ party} & \shortstack{\\ Easy \\ content \\ parsing} & \shortstack{\\ Required \\ changes \\ on} \\ \hline
    \multicolumn{1}{|l|}{TLSNotary~\cite{TLSnotary}} & --- & --- & TLS Protocol \\
    \multicolumn{1}{|l|}{TLS-N~\cite{ritzdorftls}} & \checkmark  & --- & TLS Protocol \\
    \multicolumn{1}{|l|}{Town Crier~\cite{zhang2016town}} & --- & \checkmark  & --- \\
    \multicolumn{1}{|l|}{\name} & \checkmark &  \checkmark & App \\ \hline
    \end{tabular}
    \caption{Comparison to most related works.}
    \label{tab:related:comparison}
    \vspace*{-0.6cm}
\end{wraptable}

In \autoref{tab:related:comparison} we compare \name with the competing schemes.
As shown, \name makes data feeds authentic and easy to parse without major
changes.  It is easy to implement, and it does not require modifications beyond
adding new functionalities in the content provider web service. It is an
advantages compared to the  solutions which require changes on the TLS protocol
for operating. Additionally, \name does not require an additional trusted party 
besides the content provider itself.

Moreover, we believe that the adoption of \name is much more likely than the
adoption of competing schemes. In contrast to transport-layer authentication
systems, \name requires changes only on the application layer. It also does not
require trusted hardware or relies on ubiquitous TLS certificates following
natural for HTTPS trust relationships. Last but not least, content providers are
motivated by economic incentives as \name allows them to be paid for
authenticating content which usually they publish for free.

%% file: sec/conclusions.tex
In this paper, we proposed \name, a practical system that provides authenticated
data feeds for smart contracts. In contrast to the previous work, \name
seamlessly integrates content providers with the blockchain platform. This
combination provides multiple benefits like efficient and easy data verification
without any new trusted parties, and new interesting features that the previous
platforms do not provide. Thanks to the deployed tamper-evident data structure
(\teds) that is monitored by a smart contract, content providers cannot
equivocate. To mitigate censorship, our scheme provides a blockchain based API
for querying content providers. Besides that, native to blockchain platforms
monetary transfers allow content providers to explore new business models,
where relying contracts would pay a fee for the content verification. Last
but not least, \name can be easily deployed today in the application layer
without any modifications to underlying protocols.

We plan to investigate \name and its components in other applications. One
particularly interesting example is a non-equivocation scheme for lightweight
clients. Due to placing validation logic in smart contracts, it should be more
efficient than, for instance, Catena~\cite{tomescu2017catena}, where clients
have to collect and validate all related transactions by themselves. We believe
\name could achieve the same property with much shorter proofs.

%% file: sec/appendix.tex
\section{Extended Security Discussion}
\label{sec:appendix:security}

\subsection{Data Authentication}
Our first claim is that \textit{an adversary cannot create a content on behalf
of a content provider}. To achieve that, the adversary need to either:
\begin{enumerate*}
    \item tamper authenticated proofs generated by the content providers, or
    \item update the authoritative contract on behalf of the content provider, or
    \item forge the manifest binding the authoritative contract and identity of 
    the content provider.
\end{enumerate*}
All these attacks are out of scope our adversary model.

The first attack is infeasible due to the security of the tamper-evident data
structured used~\cite{crosby2009efficient}. More specifically, generating a
membership proof for a non-element of the data structure is equivalent to
breaking a deployed hash function. Therefore, the adversary to create such a
proof for a malicious element has to extend the data structure by adding the
element and updating the authoritative contract by a new root. However, in this
attack, the adversary cannot update the authoritative contract as it
enforces the update procedure (see \autoref{sec:details:content_contract:update}). 
The update procedure allows only the contract's owner to update it. Therefore, 
without the content provider's blockchain key, the adversary cannot update the 
legitimate authoritative contract and prove on the malicious content.

For the last attack, the manifest's digital signature is verified using the TLS
PKI. Thus, without the ability to a) use a TLS private key of the content
provider, or b) obtain a digital certificate of the content provider, the
adversary cannot create a malicious manifest on behalf of the content provider.
These attacks are out of the scope of our adversary model, but we discuss them
and their implications in the next section.

\subsection{51\%-Blockchain Attack}
\label{sec:analysis:blockchain}
In this section we discuss how adversaries able to undermine the blockchain
properties (although they are outside our adversary model) can impact \name. In
particular, we focus on the 51\%-attack~\cite{nakamoto2008bitcoin} where an
adversary possesses more than 50\% of the total mining power of the blockchain
network, which would allow her to rewrite the blockchain history. Such an
adversary, could attack availability of \name (and any other blockchain
application) by reverting or denying arbitrary transactions (or even authoritative
contract creations).

An interesting scenario is an adversary colluding with a content provider.
Besides availability attacks, the adversary could allow the content provider to
equivocate by creating two conflicting \teds versions. One version would be
maintained on the ``main'' blockchain, while the second one would exist only on the
``malicious'' blockchain mined by the adversary. Such an attack violates the desired
property of keeping content providers consistent, and enables attacks similar as
double-spending attacks~\cite{karame2015misbehavior}.

Another interesting scenario is an adversary colluding with one of the contract
parties to attack another contract party.  Such an adversary cannot forge data
entries or an outcome of the membership verification. However, it is a common
practice that smart contracts define a timeout for inaction, after which deposits
of the contract parties are sent back to them. In that case, the adversary could
reverse a genuine transaction of the victim, causing the timeout from which the
colluding party would benefit.

\subsection{General Discussion}
\input{sec/discussion}

\section{Case Study and Implementation Details}

\subsection{Case Study}
\label{sec:appendix:casestudy}
In our proof of concept, we considered a scenario where contract parties decide
to settle gambling agreement creating and deploying a smart contract which uses 
trusted data from a content provider who adopts \name in its service.

\textbf{Content Provider} Following specifications in~\autoref{sec:protocol} 
and templates provided in~\autoref{appendix:contracts_and_json}, our 
implementation of the content provider is a web service which offers data of 
football matches in JSON format. We configured it to support HTTPS, and we 
obtained a free dataset from \url{https://www.football-data.org/}. We 
implemented the \teds using Keccak-256~\cite{bertoni2009keccak} as a 
cryptographic hash function. We chose Keccak as it is a state-of-the-art hash 
function (the current standard SHA-3~\cite{dworkin2015sha} is an instance of 
Keccak) and it allows us to reduce the cost of membership and consistency 
verifications due to its native support in the Ethereum platform.

\textbf{Contract parties} It is an HTTP client application able to interact with
the content provider and a relying contract. It is capable to get and validate
the authenticity of the manifest, and it is able to submit data obtained from 
the content provider to the relying contract which executes the membership 
verification, interacting with the authoritative contract, and proceeds to 
parse the JSON data. In this case, we use a JSON parser coded in Solidity since
it is not supported natively in Ethereum platform.

\subsection{Implementations}
\label{appendix:contracts_and_json}
In this section, we show examples of how JSON data look like in our 
implementation and experiments. The JSON examples are related to the case study
explained in~\autoref{sec:appendix:casestudy}.

\begin{multicols}{2}
  \begin{lstlisting}[style={lststyle}, frame=single, caption={A manifest example.}, 
    label={lst:manifest}]
  {
  "signed":{
    "url":"https://example.com/soccer",
    "sc_address":"0x539c94cb89E127...",
    "sc_interface":
      "[{"constant":true,
          "inputs":[{"name":"json",
                    "type":"string"}],
          "name":"parseJSONdata",
          "outputs":[{"name":"",
                      "type":"bool"}],
          ...}]",
    "data_structure":
      "{id:string, local:string,
        visitor:string, localGoals:int,
        visitorGoals:int}"
  },
  "signature":"63cc6a76fd07252ff4af4c..."
  }
  \end{lstlisting}
  \columnbreak
  \begin{lstlisting}[style={lststyle}, frame=single, caption={A \name data entry 
    example. It consist of the data content itself and its membership proof 
    which is an array of elements containing a hash value and a side (0 
    indicates left side and 1 indicates right one).}, label={lst:data_and_proof}]
  {
  "content":{
    "id":"341576",
    "date":"2018-07-15T18:00:00Z"
    "local":"France",
    "visitor":"Croatia",
    "localGoals":4,
    "visitorGoals":2
  },
  "proofs":[
    {"side":0, "hash":"5e41f..."},
    {"side":1, "hash":"01950..."},
    ... more items]
  }
  \end{lstlisting}
\end{multicols}

%% file: sec/discussion.tex
By analyzing the implications and costs of adopting it, we present PDFS as a viable 
alternative for smart contracts to receive authenticated data from content providers. 
In this paper, we focus on design a system with desired properties explained 
in~\autoref{sec:pre:properties}. However, we are aware of issues and disagreements 
that one might argue against our proposed solution.

Firstly, one might claim that signature verification solutions would requires less 
effort for contract providers, and further, it provides properties of authenticity 
and provenance of data. Nevetherless, as observed 
in~\autoref{sec:implementation:evaluation}, \name is cheaper regarding \emph{gas} 
cost and extends security properties to include accountability and non-equivocation 
for content providers. On the other hand, a naive solution would be to publish data 
hashes itself in a smart contract, however, that would be prohibitively expensive 
due to smart contract storage fees.

Secondly, we aimed a design for smart contracts data feed that avoids the complexity 
of alternative solutions and related works. We consider that modifying a protocol 
extensively used or including special hardware and network specifications makes a
solution highly difficult to deploy; such as modifying the TLS protocol or including
oracles using SGX. By contrast, \name offers as a simplier alternative that only 
requires changes on the application layer for content providers and contract-to-contract 
communication for contract parties. We consider it makes \name more practical and easy 
to adopt, even without taking the new business model that a content providers might 
get by providing data in a \name service.

Lastly, our current approach keeps the common trust chain with only includes contract 
parties who want to stablish an agreements and a content provider who is an autoritative 
entity who defines trustworthy data, also known as \emph{the truth}. Although the 
content provider may be able to misbahave, \name is not able to detect such actions 
due to data content is not analyzed, but that issue also affects the related 
works. However, it can be solved by including agreement protocols. For instance, the 
relying contract might revoke any agreement if two conflicting data are submitted 
within a time gap.